\documentstyle[preprint,eqsecnum,epsf,aps]{revtex}

\begin{document}
\draft \title{The Low Energy Limit of the Chern--Simons Theory Coupled
  to Fermions\cite{byline}}
\author{H. O. Girotti$^a$, M. Gomes$^b$, J. R. S. Nascimento$^b$ and A. J.
  da Silva$^b$} \address{$^a$ Instituto de F\'{\i}sica, Universidade
  Federal do Rio Grande do Sul, CP15051, 91501-970 Porto Alegre, RS, Brazil.\\
  $^b$ Instituto de F\'\i sica, Universidade de S\~ao Paulo, CP66318,
  05315--970, S\~ao Paulo, SP, Brazil.}

\date {March 24, 1997} \maketitle

\begin{abstract}
We study the nonrelativistic limit of the theory of a quantum
Chern--Simons field minimally coupled to Dirac fermions. To get the
nonrelativistic effective Lagrangian one has to incorporate vacuum
polarization and anomalous magnetic moment effects. Besides that, an 
unsuspected quartic fermionic interaction may also be induced. 
As a by product, the method we use to calculate loop diagrams, separating
low and high loop momenta contributions, allows to identify how a   quantum
nonrelativistic theory nests in a relativistic one. 
\end{abstract}

\narrowtext

\section{Introduction}

Nonrelativistic quantum field theories are important to the
description and clarification of conceptual aspects of the physics of
systems in the low energy regime. One example of such situation is
provided by the treatment of the Aharonov Bohm effect \cite{Ah} by means of the
model of a scalar field minimally coupled to a Chern--Simons (CS) field. In
order to achieve accordance between the exact and perturbative one
loop calculations, it was necessary to include in the perturbative
approach a quartic scalar self interaction with a coupling tuned to
eliminate divergences and to restore the conformal invariance of the
tree approximation \cite{Lo}. It was later also shown that this
quadrilinear interaction automatically arises in the low energy limit
of the corresponding full-fledged relativistic quantum theory
\cite{Bo}.

In this paper we will extend the above considerations to the fermionic
case.  Specifically, we study the low energy limit of the 2+1
dimensional theory of a CS field minimally coupled to
fermions, as specified by the Lagrangian density\cite{Ha}

\begin{equation}
\label{1}
{\cal L} = \frac{\theta}{4}
\epsilon^{\mu\nu\alpha}F_{\mu\nu}A_{\alpha}
+\frac{i}{2}\bar{\psi}\gamma^{\mu}\partial_{\mu}\psi-
\frac{i}{2}(\partial_{\mu}\bar{\psi})\gamma^{\mu}\psi-m\bar{\psi}\psi
+e \bar \psi\gamma^\mu \psi A_\mu,
\end{equation}

\noindent
where $F_{\mu\nu} = \partial_\mu A_\nu - \partial_\nu A_\mu$ and
$\psi$ is a two component Dirac field representing a fermion and an
anti-fermion of the same spin \cite{f1}.  In particular we investigate
up to which extent the Pauli--Schr\"odinger (PS) nonrelativistic
Lagrangian

\begin{eqnarray}
\label{2}
{\cal L} &=& \phi^\dagger (i\frac{d\phantom{x}}{dt}- e A^0) \phi -
\frac{1}{2 m} (\vec \nabla\phi - i e \vec A \phi)^\dagger\cdot (\vec
\nabla\phi - i e \vec A \phi) \nonumber \\
&\phantom& +\frac{e}{2m} B
\phi^\dagger \phi + \frac{\theta}{4} \epsilon_{\mu\nu\rho} A^\mu
F^{\nu\rho},
\end{eqnarray}

\noindent 
where $\phi$ is a one--component anticommuting Pauli
field, correctly describes the low energy limit of (\ref{1}). Here $B
= -F_{12}$ is the magnetic field. 

By rescaling the $A^\mu$ field we could formulate the models above with
only one dimensionless interaction strength $(e/\sqrt {\rm \theta})$.
Actually, this is the small parameter which appears in our
perturbative expansions. Nevertheless, (\ref{1}) and (\ref{2}) are
conventional in the literature, $A_\mu$ having the same canonical dimension as
a Maxwell field in $2+1$ dimension.

To fix ideas, we choose once and for all to work in the
Coulomb gauge, where the $A_\mu$ free propagator is known to be

\begin{equation}
\Delta_{\mu\nu}(k)= \frac{1}{\theta}\epsilon_{\mu\nu\rho}
\frac{\bar k^\rho}{\vec k^2},\label{3}
\end{equation}

\noindent
where $\bar k^\alpha \equiv (0,\vec k)$. The $k^0-$independence of
this propagator is the basic reason why the use of the Coulomb gauge
is convenient for nonrelativistic calculations.

Let us begin by pointing out that  (\ref{2}) can not fully reproduce the
two and three point vertex functions arising from (\ref{1}). This is
due to the absence of one loop radiative corrections in (\ref{2}),
which follows from the antisymmetry of (\ref{3}) and the fact that
nonrelativistic fermions only propagate forward in time.  On the other
hand, the radiative corrections arising from (\ref{1}) do not
vanish. The contributions to the two and three point functions are,
actually, superficially divergent and need to be subtracted.

We begin our study by investigating the nonrelativistic model
(\ref{2}). In section II we verify the absence of one loop radiative
corrections to the two and three point vertex functions. We study also
the fermion--fermion scattering amplitude and confirm the assertion
made in \cite{Lo} that the Pauli's interaction term regularizes the
theory and gives a contribution essential to reproduce the correct
Aharonov--Bohm amplitude up to order $e^4$.

In section III we examine the relativistic model (\ref{1}). After
summarizing the renormalization program \cite{Gi} we prove that
radiative corrections induce both an ``anomalous'' magnetic term and a
Maxwell term, absents in (\ref{2}).  To calculate the one loop
corrections to the fermion--fermion scattering amplitudes we
will employ a scheme which separates the contributions of the low and
high momenta intermediary states \cite{Go}. This allows a direct
simplification of the integrands and it is closely related to the
methods of effective field theories \cite{Wi}. We will show that the
low momenta intermediary part coincides with the result for the same
process calculated from the Lagrangian (\ref{2}). On the other hand,
the contributions from the high intermediary momenta can be thought as
coming from new interactions in the effective nonrelativistic theory.
We finish this section presenting a discussion of our results.

\section{Nonrelativistic Theory}
Our purpose here is to use (\ref{2}) to study the interaction of the
CS field with fermions up to the one loop approximation. Afterwards,
we shall compare these results with those for Dirac fermions,
calculated from (\ref{1}) in the low energy regime.

As said in the introduction, we shall work in the radiation gauge where the
free gauge field propagator is given in (\ref{3}).
From that expression we can get the free propagator
\begin{equation}
\Delta_B(x)=\langle {\rm T} B(x) A^0(0) \rangle = -\frac{i}{\theta} \delta^3(x)\label{4}
\end{equation}
which is also necessary to construct Feynman amplitudes.

As the second quantized free fermionic field is
\begin{equation}
\phi(\vec x, t) = \int \frac{d^2k}{2 \pi}\, b(\vec k)\,\, {\rm e}^{-i(
\frac{{\vec k}^{ 2}}{2m}t- \vec k \,\, \cdot \vec x)},
\end{equation}
where the annihilation operator $b(\vec k)$ satisfies
\begin{equation}
\{b(\vec k),\, b^\dagger ({\vec k}^\prime)\}= \delta^2(\vec k -
{\vec k}^\prime),
\end{equation}
the free $\phi$--propagator turns out to be given  by
\begin{eqnarray}
G(x-y)&\equiv &\langle {\rm T} \phi(x) \phi^\dagger(y)\rangle \nonumber \\
&=& \theta(x^0-y^0) \langle \phi(x) \phi^\dagger(y)\rangle =\int \frac{d^3k}
{(2\pi)^3} {\rm e}^{-ik^0(x^0 -y^0-\eta) + i \vec k \cdot (\vec x - \vec y)} 
\frac{i}{k^0-\frac{{\vec k}^2}{2 m} + i \epsilon},\label{41}
\end{eqnarray}
where $\eta$ and $\epsilon$ are positive parameters to be set to zero
at the end of the calculations.  In the above definition a time
ordering prescription was chosen so that the propagator is zero for
$x^0=y^0$. This implies that any closed fermion loop automatically
vanishes.

\begin{figure}
\centerline{\epsfbox{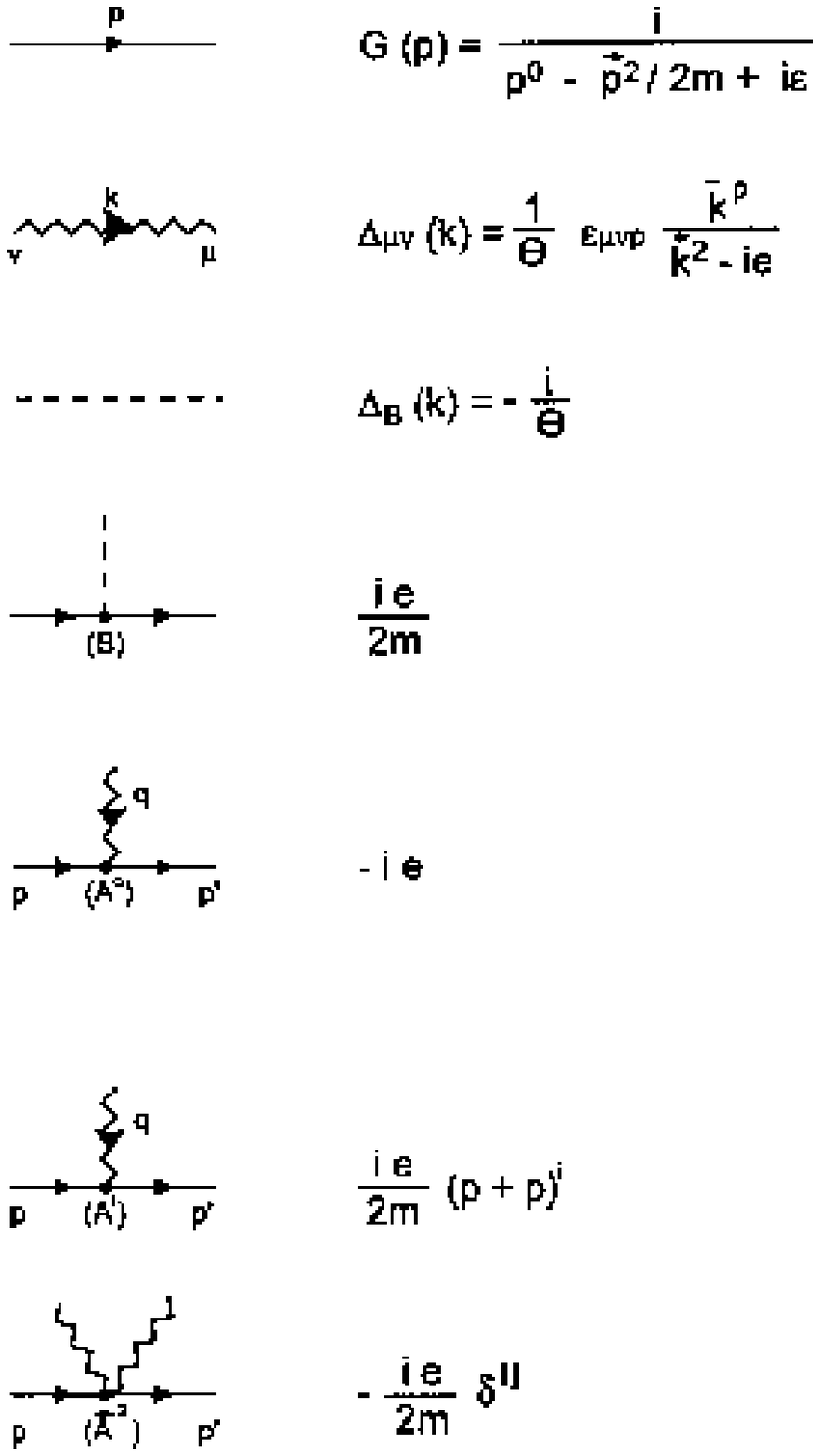}}
\caption{Feynman rules for the PS theory.} 
\label{fig1}
\end{figure}

\begin{figure} 
\centerline{\epsfbox{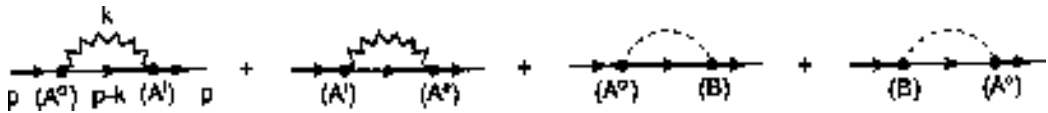}}
\caption{One loop contributions to the PS fermion self energy.}
\label{fig2}
\end{figure}

Our graphical notation 
is shown in Fig.\ \ref{fig1}. Using these rules, one can
demonstrate that there are no one loop corrections to the propagators
and vertices. The would be one loop corrections to the fermion two
point function are represented in Fig.\ \ref{fig2}. The first two of those
diagrams cancel each other due to the anti-symmetry of the $A_\mu$
propagator and the third and fourth are in fact closed fermionic loops
(see (\ref{4})) and therefore give no contribution as remarked after
equation (\ref{41}).  Similar arguments allows to extend these
conclusions to all remaining one loop vertex and propagators graphs.

We will next study the fermion--fermion elastic scattering.
To consider the possibility of the scattering of non identical
fermions we will not anti-symmetrize our amplitudes. The incoming and
outgoing fermions are assumed to have momenta $p_1= (\frac{{\vec
    p_1}^{\phantom{x}2}}{2m}, \vec p_1)$, $p_2= (\frac{{\vec
    p_2}^{\phantom{x}2}}{2m}, \vec p_2)$ and $p_{1}^{\prime}=
(\frac{\vec p_{1}^{\phantom{x}\prime 2}}{2m}, \vec
p_{1}^{\phantom{x}\prime})$, $p_{2}^{\prime}= (\frac{\vec
  p_{2}^{\phantom{x}\prime 2}}{2m}, \vec p_{2}^{\phantom{x}\prime})$,
respectively. We shall work in the center of mass frame where $\vec p_1 =
-\vec p_2 =\vec p$, $\vec p_{1}^{\phantom{x}\prime} = -\vec
p_{2}^{\phantom{x}\prime} =\vec p^{\phantom{x}\prime}$ and $|\vec p|=
|\vec p^{\phantom{x}\prime}|$. 

\begin{figure} 
\centerline{\epsfbox{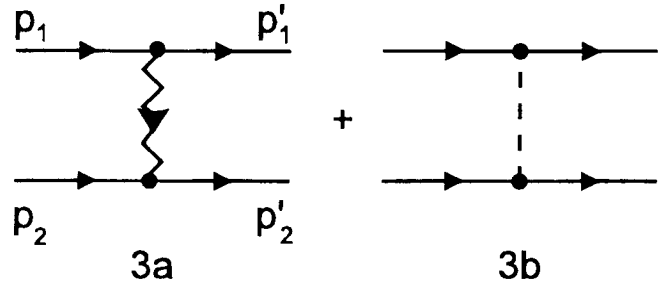}} 
\caption{Graphs for the PS fermion-fermion
  scattering in the  tree approximation.} \label{fig3} 
\end{figure}

Up to one loop the non vanishing contributions come from the diagrams
in Figures\ \ref{fig3} and \ \ref{fig4}.  The amplitude corresponding
to the diagrams in Fig.\ \ref{fig3} is
\begin{equation}
A^{(0)}= \frac{e^2}{m\theta}\Bigl [ 1+ i \frac{\vec s \times \vec
q}{\vec q^2}\Bigr ],\label{5}
\end{equation}
\noindent
where $\vec s=\vec p + \vec p^{\phantom{x}\prime}$, $\vec q = \vec
p^{\phantom{x}\prime}-\vec p$ and $\vec s \times \vec q$ stands for
$\epsilon^{ij}s_i q_j$. The $\vec q$ dependent term result from the
graph containing the propagator $\langle {\rm T} A^0 A^i\rangle$ and
the $\vec q$ independent one comes from the contact Pauli interaction
mediated by $\langle {\rm T}B A^0\rangle$.

Let us now examine the one loop diagrams. We will take the procedure of
first doing the $k^0$ integration and then regularizing the remaining $\vec k$
integration by a cutoff $\Lambda$.
The box
diagrams \ \ref{fig4}a yield

\begin{eqnarray}
A_{box}^{(1)}&=&-\frac{e^4 i}{m^2 \theta^2} \int_{0}^{\Lambda}
\frac{d^3k}{(2 \pi)^3}
\frac{1}{k^0+ \frac{\vec p^2 - \vec k^2}{2m}+
i\epsilon}\,\,\frac{1}{-k^0+ \frac{\vec p^2 - \vec k^2}
{2m}+i\epsilon}\nonumber\\
&\phantom{x}&\epsilon_{ij}\epsilon_{ln}\frac{( k-{
p}^{\phantom{x}\prime})^j}{|\vec k-\vec
p^{\phantom{x}\prime}|^2}\frac{( k-{ p})^n}{|\vec k-\vec
p|^2} (- p^{\phantom{x}\prime} - k)^i(
p^{\phantom{x}\prime} + k)^l\nonumber \\[16pt] 
&=& -\frac{4 
e^4}{m \theta^2} \int_{0}^{\Lambda} \frac{d^2k}{(2 \pi)^2}
\frac{1}{\vec k^2-\vec p^2-i\epsilon}
\frac{\vec p^{\phantom{x}\prime} \times \vec k}{|\vec k-\vec
p^{\phantom{x}\prime}|^2}
\frac{\vec p \times \vec k}{|\vec k-\vec p|^2}\nonumber\\[16pt]
&=&\frac{ e^4}{4 \pi m \theta^2}\Bigl \{ \ln[\frac{{\vec q}^2}{{\vec p}^2}]
+ i\pi\Bigr  \}
+ {\cal O}({\vec p}^2/\Lambda^2),
\end{eqnarray}
\noindent
while the ``contact'' diagram\ \ref{fig4}b gives
\begin{eqnarray}
A_{cont}^{(1)}&=&i\frac{e^4}{m^2 \theta^2} \int_{0}^{\Lambda} \frac{d^3k}
{(2 \pi)^3}
\frac{1}{k^0+ \frac{\vec p^2 - \vec k^2}{2 m}+
i\epsilon}\,\,\frac{1}{-k^0+ \frac{\vec p^2 - \vec k^2}{2 m}+
i\epsilon}\nonumber \\[16pt] 
&=& - \frac{e^4}{m \theta^2} \int_{0}^{\Lambda}
\frac{d^2k}{(2 \pi)^2}\frac{1}{{\vec k}^2 - {\vec p}^2 - i
\epsilon} = - \frac{e^4}{4 \pi m \theta^2}\{\ln\frac{\Lambda^2}{{\vec p}^2}
+i \pi\} +{\cal O}({\vec p}^2/\Lambda^2).
\end{eqnarray} 

\noindent
Finally, the triangle graphs 4c and 4d give
\begin{eqnarray}
A_{tri}^{(1)}&=&-\frac{2e^4 i}{m \theta^2} \int_{0}^{\Lambda}
\frac{d^3k}{(2 \pi)^3}
\frac{1}{-k^0+ \frac{\vec p^2}{2m} -\frac{|\vec p+ \vec k|^2}{2 m}+i\epsilon}
\epsilon^{il}\epsilon^{in}
\frac{ k^l}{|\vec k|^2}\frac{( p^{\phantom{x}\prime}- p-k)^n}
{|\vec p^{\phantom{x}\prime} - \vec p -\vec k|^2}
\nonumber \\[16pt]
&=& -\frac{ e^4 }{m \theta^2} \int_{0}^{\Lambda} \frac{d^2k}{(2 \pi)^2}
\frac{\vec k \cdot (\vec p^{\phantom{x}\prime} - \vec p- \vec k)}
{|\vec k|^2 |\vec p^{\phantom{x}\prime} - \vec p -\vec k |^2}, \label{17}
\end{eqnarray}
where the $k_o$ integration was done symmetrically. The final result is then
\begin{equation}
A_{tri}^{(1)}=- \frac{e^4}{4 \pi m \theta^2}\{ \ln\frac{{\vec q}^2}
{\Lambda^2}\}+{\cal O}({\vec p}^2/\Lambda^2).\label{171}
\end{equation}
By summing up the above contributions we get 
\begin{eqnarray}
A &=& A^{(0)}+A_{cont}^{(1)} + A_{box}^{(1)}+A_{tri}^{(1)}=  
\frac{e^2}{m\theta}\Bigl [ 1+ i \frac{\vec s \times \vec q}{\vec q^2}\Bigr ]
+{\cal O}({\vec p}^2/\Lambda^2)\nonumber \\
&\phantom{x}&\label{172}
\end{eqnarray}

\noindent
This result shows that, up to one loop, there is no radiative
correction to the nonrelativistic scattering. This holds for all
values of the coupling constant $e$. In the model of a nonrelativistic
boson coupled to a CS field, a similar result was first obtained in
\cite{Lo}. There the role of the contact Pauli interaction was
played by a $\lambda (\phi^\dagger \phi)^2$ interaction with $\lambda$
chosen to restore the scale invariance \cite{Ja} present in the tree
approximation. Observe also that in terms of the angle $\chi$
between $\vec p$ and ${\vec p}^{\phantom{x} \prime}$, equation
(\ref{172}) becomes
\begin{equation}
 i \frac{e^2}{m \theta} \frac{ {\rm e}^{-i\chi/2}}{\sin(\,\chi/2)}
\end{equation}
which is the expansion up to order $e^4$ of the Aharonov-Bohm amplitude for
fermions \cite{Pe}.
\section{Relativistic Theory}

We now consider the relativistic theory defined by (\ref{1}). The
corresponding Feynman rules are depicted in Fig.\ \ref{fig5}. By power counting
the model is renormalizable, the degree of divergence of a graph
$\gamma$ being $\delta(\gamma)=3-F -B$, where $F$ and $B$ are the
number of external fermion and boson lines, respectively. Thus, the
only divergences are those associated to the fermion two point
function, the CS two point function and to the vertex. The
renormalization of the model in the Coulomb gauge, up to one loop, was
studied in \cite{Gi}, using dimensional regularization. Here, for
completeness, we just stress the main points of that calculation.

 The ambiguities in the finite parts
are eliminated by adding to (\ref{1}) the counterterm Lagrangian
density

\begin{equation}\label{31}
{\cal L}_{c}= \delta Z\,(\frac{i}{2}\bar{\psi}\gamma^{\mu}\partial_{\mu}\psi-
\frac{i}{2}(\partial_{\mu}\bar{\psi})\gamma^{\mu}\psi) - \delta m \,\, 
\bar{\psi}\psi + \frac{\delta\theta}{4}
\epsilon^{\mu\nu\alpha}F_{\mu\nu}A_{\alpha} +\delta e\,\, \bar{\psi}
\gamma^{\mu}A_{\mu}\psi
\end{equation}
where the coefficients $\delta Z$, $\delta m$, $\delta \theta$ and
$\delta e$ are fixed by the normalization conditions specifying the
$\psi$ field intensity, the values of the physical mass, CS
parameter and charge, respectively.

First, consider CS self energy, 
\begin{eqnarray}
\Pi^{\alpha \beta}(k)\,&=& \, -e^2\int\frac{d^3q}{(2\pi)^3}\frac{{\rm
Tr}[\gamma^\alpha(\not \!q+\not \!k+m)\gamma^\beta(\not \!q
+m)]}{[(k+q)^2-m^2](q^2-m^2)} - 2 \delta \theta \epsilon^{\alpha\beta\rho}
k_\rho\nonumber\\[16pt]
&=& \,- i\,\,(k^2\,g^{\alpha \beta} \,-\,k^{\alpha}
 k^{\beta})\,A(k^2))\,+\,\epsilon^{\alpha \beta \rho}\,k_{\rho}\,(B(k^2)-2 
\delta \theta)\,\,,
\label{12}
\end{eqnarray}

where 

\begin{mathletters}
\begin{equation}
A =  \left(\frac{m}{k^2} + \frac{1}{4m} \right) \,B(k^2) \,
- \,\frac{e^2 m}{4 \pi k^2}\,\,\,,\label{mlett:a54} \\
\end{equation}
\end{mathletters}
and

\begin{equation}
\label{55}
B(k^2)\,=\frac{e^2 m}{4 \pi}\,\,\int_{0}^{1} dx \,\frac{1}{\sqrt{k^2x (x-1)
+m^2-i\epsilon}}\, .
\end{equation}

\noindent
By choosing $\delta\theta= B(k=0)/2= \frac{e^2}{8 \pi}\varepsilon(m)$,
where $\varepsilon$ denotes the sign function, we fix $\theta$ to be
the renormalized CS parameter (this renormalization could,
equivalently, be interpreted as a wave function renormalization for
$A_\mu$). For low momentum, $\Pi^{\alpha\beta}$ approaches the
expression

\begin{equation}
\label{551}
\Pi^{\alpha\beta}(k)= - i\frac {e^2}{12 \pi |m|}(k^2 g^{\alpha\beta}- k^\alpha 
k^\beta), 
\end{equation}
showing the well known phenomena of induction of a Maxwell term in the
effective Lagrangian of the model \cite{De}.

The fermion self energy is
\begin{eqnarray}
\label{501}
\Sigma(p) &=& -\frac {e^2 i}{\theta} \int \frac {d^3k}{(2 \pi)^3}
\frac{\gamma^\mu (\not \!k+ \not \!p +m)\gamma^\nu}{(k+p)^2-m^2}
\frac {\epsilon_{\mu\nu\rho}{\bar k}^{\rho}}{({\vec k})^2}+ i \delta Z 
\, \not \!p - i \delta m \nonumber \\
&=& \, \frac {i e^2}{ 2 \pi \theta}\Bigl [\frac {({\vec p}^2- m \vec p 
\cdot \vec \gamma)}{m+w_p} - \, m \Bigr ]+ i \delta Z \, \not \!p - i \delta m .
\end{eqnarray}
Choosing $\delta m  \,=\,- \frac{m e^2}{2\pi \theta}$, we guarantee that the
pole of the fermion propagator up to this order is at $p^2 = m^2$. Besides
that, taking $\delta Z=0$ the form of the propagator in the fermion rest
frame is the same as for the free case \cite{Ad}. 

Finally, the  radiative correction to the vertex is given by
\begin{eqnarray}
\label{508}
\Gamma^\rho(p, p') &=& 
\frac {i e^3}{\theta} \int \frac {d^3k}{(2 \pi)^3}
\phantom {abcdfghojklmnopqrstuvx}\nonumber \\
&& \frac{\gamma^\mu (\not \!p' - \not \!k+m)\gamma^\rho (\not \!p - 
\not \!k+m)\gamma^\alpha \epsilon_{\alpha \mu \nu}{\bar k}^\nu }{[(p'-k)^2 
- m^2+i \epsilon][(p-k)^2 - m^2+i \epsilon](-{\vec k}^2)}+
i(e + \delta e)\gamma^\rho .\nonumber\\
&&
\end{eqnarray}  

\noindent
Thus, choosing $\delta e=0$, we get in the low momentum regime
\begin{eqnarray}
\label{517}
\bar{u}(p') \Gamma^0 u(p) &=& ie, \qquad \\
\bar{u}(p')\Gamma^i u(p) &=& i e\left (1 - \frac{e^2}{4\pi\theta}\right ) 
\frac{(p+p'^i)}{2 m} + e \left(
1+\frac{e^2}{4 \pi \theta}\right ) \epsilon^{ij}\frac{(p'-p)^j}{2m} .
\end{eqnarray}

In a covariant gauge the magnetic moment of the fermion could be read
as the coefficient of $\epsilon^{ij}(p'-p)^j$ in this last expression.
This happens because only the first of the three diagrams of Fig.\ \ref{fig6},
which appear in the calculation of the scattering of the fermion by an
external field $\cal A^\rho$, is nonvanishing on shell.  In the
Coulomb gauge this is not so and only after taking into account the
contribution of all three diagrams we get $\frac{e^3}{4\pi m\theta}$
for the anomalous magnetic moment of the fermion. This result is in
accord with calculations in covariant gauges where only graph 6a
contributes \cite{Se}.

It is now clear that, up to one loop, instead of (\ref{2}) these
radiative corrections induce the following nonrelativistic Lagrangian,

\begin{eqnarray}
\label{21}
{\cal L}_{eff} &=& \phi^\dagger (i\frac{d\phantom{x}}{dt}- e A^0) \phi
- \frac{1}{2 m} (\vec \nabla\phi - i e \vec A \phi)^\dagger\cdot (\vec
\nabla\phi - i e \vec A \phi) \nonumber \\ &\phantom& +\frac{e}{2m}\,g \, B
\phi^\dagger \phi + \frac{\theta}{4} \epsilon_{\mu\nu\rho} A^\mu
F^{\nu\rho} - \frac{1}{4}\left (\frac{e^2}{12 \pi m}\right
)F^{\mu\nu}F_{\mu\nu},
\end{eqnarray}

\noindent
where $g \equiv 1+ e^2/2\pi\theta$. 

We shall next look for the appearance of a $(\phi^\dagger \phi)^2$
vertex in the effective nonrelativistic Lagrangian.  We so focus on
the elastic fermion--fermion scattering amplitude.  In the center of
mass frame, the incoming and outgoing fermions are assumed to have
momenta $p_1=(w_p,\vec p)$, $p_2= (w_p,-\vec p)$ and
$p_{1}^{\phantom{x}\prime}= (w_p,\vec p^{\phantom{x} \prime})$,
$p_{2}^{\phantom{x}\prime}= (w_p,-\vec p^{\phantom{x} \prime})$, where
$|\vec p|= |{\vec p}^{\phantom {x}\prime}|$ and $w_p= \sqrt{m^2+ {\vec p}^2}$.

The various contributions, up to one loop, are shown in Figs.\ 
\ref{fig7} and \ \ref{fig8}.  The tree approximation (graph \ \ref{fig7}) is
given by:
\begin{equation}
T^{(0)}= -i e^2 \bar u(\vec {p}^{\phantom {x}\prime})\gamma^\mu u(\vec {p})
\Delta_{\mu\nu}(\vec {p}^{\phantom {x}\prime}-\vec {p}) \bar 
u(-\vec {p}^{\phantom {x}\prime})\gamma^\nu u(-\vec {p})
\end{equation}
Its low energy approximation is get by expanding $w_p=
(m^2+p^2)^{1/2}$ in powers of $\frac{|\vec p|}{m}\; (\ll 1)$. To leading
order, we have:
\begin{equation}
\label{6}
T^{(0)}= \frac {e^2}{\theta m} (1 + i\frac{\vec s \wedge \vec q}
{{\vec q}^2}). \label{f15}\end{equation}

Observe that (\ref{f15}) is the same as the $e^2$-amplitude (\ref{5}) in the
PS theory, due to exchange of one photon, including the contribution
of the Pauli interaction.

Self-energy and vertex radiative corrections to the tree approximation (Fig.\ 
\ref{fig7}), in leading $1/m$ order, give
\begin{equation}
T_R=\frac{ e^4}{12 \pi m\theta^2}+\frac{ e^4}{2 \pi m\theta^2}= \frac{7 e^4}{12 \pi m\theta^2},
\end{equation}
where the first and second terms in the first equality come, respectively, from the
vacuum polarization and vertex insertions. 
$T_R$ must not be considered for the induction of a term
$(\phi^\dagger \phi)^2$ since self-energy and vertex corrections have already
been 
incorporated in (\ref{21})through  the fermion anomalous magnetic moment and the Maxwell terms. 

It remains to calculate the graphs in Fig.\ \ref{fig8}a and \ \ref{fig8}b.
They are
respectively given by (the subscripts $B$ and $X$ stand for box and
crisscross two photons exchange amplitudes)

\begin{eqnarray}
\label{8}
&& T_B =i e^4 \int \frac {d^3 k}{(2 \pi)^3} \Bigl [\bar 
u(\vec {p}^{\phantom {x}\prime})\gamma^\mu S_F(k) \gamma^\nu 
u(\vec p)\Bigr ] 
\nonumber \\
&&\Bigl [\bar u(-\vec {p}^{\phantom {x}\prime})\gamma^\alpha S_F(-k) 
\gamma^\beta u(-\vec p)\Bigr ] \Delta_{\nu\beta}(\vec k - \vec p) 
\Delta_{\alpha\mu}(\vec k - \vec p^{\phantom {x}\prime})
\end{eqnarray}
and

\begin{eqnarray}
\label{9}
&& T_X=i e^4 \int \frac {d^3 k}{(2 \pi)^3} \Bigl [\bar 
u(\vec {p}^{\phantom {x}\prime})\gamma^\mu S_F(k) \gamma^\nu 
u(\vec p)\Bigr ] 
\nonumber \\
&&\Bigl [\bar u(-\vec {p}^{\phantom {x}\prime})\gamma^\alpha S_F(k 
-p -p^\prime) 
\gamma^\beta u(-\vec p)\Bigr ] \Delta_{\nu\alpha}(\vec k - \vec p) 
\Delta_{\beta\mu}(\vec k - \vec p^{\phantom {x}\prime}).
\end{eqnarray}

In what follows the free fermion propagator will be written in terms
of fermion and anti-fermion wave functions \cite{f1},

\begin{equation}
S_F(p)=
i\frac{u(\vec p)\bar u(\vec p)}{p^0-w_p +i\epsilon}+i\frac{v(-\vec p)\bar
v(-\vec p)}{p^0+w_p -i\epsilon}.\label{101}
\end{equation}
 
\noindent
This device greatly simplifies the calculation of the $k^0$ integrals.
As a by product we can trace the contribution of fermions or
anti-fermions in intermediary states.
  
Replacing (\ref{101}) in the expressions above, we get

\begin{eqnarray}
&& T_B=- ie^4 \int \frac {d^3 k}{(2 \pi)^3} \Delta_{\nu\beta}(
\vec k - \vec p) \Delta_{\alpha\nu}(\vec k - \vec p^{\phantom {x}\prime}) 
\nonumber \\
&&\bar u (\vec p^{\phantom {x}\prime}) \gamma^\mu \Bigl [ \frac {u(\vec k) 
\bar u (\vec k)}{k^0+ w_p-w_k + i \epsilon}+ \frac {v(-\vec k)
\bar v(-\vec k)}{k^0- w_p+w_k - i \epsilon}\Bigr ]\gamma^\nu 
u (\vec p)\nonumber \\
&& \bar u (-\vec p^{\phantom {x}\prime}) \gamma^\alpha \Bigl 
[ \frac {u(-\vec k) \bar u (-\vec k)}{w_p-k^0-w_k + i \epsilon}+ 
\frac {v(\vec k) \bar v(\vec k)}{w_p-k^0+w_k - i \epsilon}\Bigr ]
\gamma^\beta u (-\vec p)
\end{eqnarray}

and

\begin{eqnarray}
&& T_X=-i e^4 \int \frac {d^3 k}{(2 \pi)^3} \Delta_{\nu\alpha}(
\vec k - \vec p) \Delta_{\beta\mu}(\vec k - \vec p^{\phantom {x}\prime}) 
\nonumber \\
&&\bar u (\vec p^{\phantom {x}\prime}) \gamma^\mu \Bigl [ \frac {u(\vec k) 
\bar u (\vec k)}{k^0+ w_p-w_k + i \epsilon}+ \frac {v(-\vec k)\bar 
v(-\vec k)}{k^0- w_p+w_k - i \epsilon}\Bigr ]\gamma^\nu u (\vec p)
\nonumber \\
&& \bar u (-\vec p^{\phantom {x}\prime}) \gamma^\alpha \Bigl [ 
\frac {u(\vec k- \vec s) \bar u (\vec k-s)}{k^0+w_p-w_{k-s} + i \epsilon}
+ \frac {v(\vec s-\vec k) \bar v(\vec s-\vec k)}{k^0+w_p+w_{k-s} 
- i \epsilon}\Bigr ]\gamma^\beta u (-\vec p).
\end{eqnarray}

The integration in $k^0$ can be done by closing the contour in the
upper half $k^0$ complex plane. After some simplifications, we get,
\begin{equation}
T_B=T^{el,el}_{B}+ T^{pos,pos}_{B}
\end{equation}

\noindent
where

\begin{eqnarray}
&&T^{el,el}_{B}= - \frac{e^4}{2} \int \frac{d^2k}
{(2 \pi)^2}\frac {w_k+w_p}{\vec k^2-\vec p^2+i\epsilon} F^{*}
(\vec k,\vec p^{\phantom {x}\prime}) F(\vec k,\vec p)
\end{eqnarray}
\noindent
and

\begin{eqnarray}
&&T^{pos,pos}_{B}= - \frac{e^4}{2} \int \frac{d^2k}{(2 \pi)^2}
\frac{1}{w_k+w_p}H(\vec p^{\phantom {x}\prime},\vec k) H^*(\vec p,\vec k).
\end{eqnarray}

For $T_X$ it results:

\begin{eqnarray}
T_X &=& \frac{e^4}{2} \int \frac{d^2k}{(2 \pi)^2}\frac {1}
{w_k+w_{k-s}}\Bigl[ G(\vec k -\vec s,-\vec p,\vec p^{\phantom {x}\prime}) 
G^*(\vec k -\vec s,-\vec p^{\phantom {x}\prime},\vec p)+\Bigr.\nonumber\\
&& \Bigl. G(\vec k,\vec p,-\vec p^{\phantom {x}\prime}) G^*(\vec k,
\vec p^{\phantom {x}\prime},-\vec p)\Bigr],
\end{eqnarray}
\noindent
where, as before,  $\vec s= \vec p + \vec p^{\phantom {x} \prime}$, and
\begin{eqnarray}
\label{10}
&& F(\vec k, \vec p)= [\bar u(\vec {k})\gamma^\mu u(\vec {p})]
\Delta_{\mu\nu}[(\vec {k}-\vec {p}) \bar u(-\vec {k})\gamma^\nu 
u(-\vec {p})],\nonumber\\
&& H(\vec p, \vec k)=  [\bar u(\vec {p})\gamma^\mu v(-\vec {k})]
\Delta_{\mu\nu}[(\vec {p}-\vec {k}) \bar u(-\vec {p})\gamma^\nu 
v(\vec {k})],\\
&& G(\vec a, \vec b,\vec c)=[ \bar u(\vec {a})\gamma^\mu u(\vec {b})]
\Delta_{\mu\nu}[(\vec {a}-\vec {b}) \bar u(\vec {c})\gamma^\nu 
v(\vec b-\vec {a}-\vec c)].\nonumber
\end{eqnarray}

The terms $T_{B}^{el,el}$ and $T_{B}^{pos-pos}$ are,
respectively, the contributions of the $u$ and $v$ fermion
wave functions (electron and positron)to the two internal fermion  
lines in Fig.\ \ref{fig8}a.
 Mixed contributions, in which $u$ ``runs'' in one line and $v$ in
the other cancel in this model. On the other hand each of the two terms in
$T_{X}$ corresponds to a graph in which one of the
internal fermion line is a $u$ and the other a $v$.  

We will break the integration region $|\vec k|=(0,\infty)$ in two
ranges:

1- Contributions of the nonrelativistic intermediate
states,corresponding to the loop momentum in the range: $|\vec
k|=(0,\Lambda)$ where $\Lambda$ is a parameter satisfying: $|\vec
p|<<\Lambda<<m$.

2- The relativistic energy intermediate states contributions,
corresponding to the range $|\vec k|=(\Lambda,\infty)$ for the loop
momentum.  

In the region $(0,\Lambda)$ the integrands can be expanded in powers
of $1/m$ up to the desired order of approximation $(w_k = m +
\frac{\vec k^2}{2m}+\frac{\vec k^4}{8m^3}+ ...)$. We will limit
ourselves to the leading $(1/m)$ order which suffices for comparison
with the nonrelativistic PS theory. For the region $(\Lambda,
\infty)$, we will expand $w_p$ around $\vec p = 0$, but keep $w_k$
exact. So, to extract the leading $(1/m)$ approximation of this part
of the integral, an extra expansion in $1/m$ must be made after the
integral is computed.  With these mentioned approximations, we can
write (\ref{8}) - (\ref{9}) in leading order as:

\begin{eqnarray}
\label{14}
T_B^{el,el}& = & \;- \frac{e^4}{\theta^2 m}\int_{0}^{\Lambda}
\frac{d^2k}{(2\pi)^2}\; \frac{1}{{\vec k}^2 -{\vec p}^2 +i\epsilon}
\Bigl [ 1+ 4 \frac{{\vec k} \wedge {\vec p}}{({\vec k}-{\vec p})^2}
\frac{{\vec k} \wedge {\vec p^{\phantom {x}\prime}}}{({\vec k}
-{\vec p^{\phantom {x}\prime}})^2}\Bigr ]\nonumber\\
&\phantom {x}&  -\frac{e^4}{2 \theta ^2} 
 \int _{\Lambda}^{\infty} \frac{d^2k}{(2\pi)^2}\; \frac{w_k +m}{{\vec k}^2 
w_k^2}\\
\label{15}
T_B^{pos,pos}& = & -\frac{e^4}{2\theta^2 }\int_{\Lambda}^{\infty}
\frac{d^2k}{(2\pi)^2}\frac{1}{(w_k +m)w_k^2}\\
\label{13}
T_X &=& \; \frac{e^4}{ m\theta^2 }\int_{0}^{\Lambda}
\frac{d^2k}{(2\pi)^2}\; \frac{{\vec k}}{{\vec k}^2}
\cdot \frac{({\vec k}-{\vec q})}{({\vec k}-{\vec q})^2}
 +\frac{e^4 m^2}{\theta^2 }\int_{\Lambda}^{\infty}
\frac{d^2k}{(2\pi)^2}
\; \frac{1}{{\vec k}^2 w_k^3}
\end{eqnarray}

In the integration region $(0,\Lambda)$ of $T_B^{el,el}$ the
contributions of graphs in which one of the photon propagators is
$<{\rm T}A^0A^i>$ and the other $<{\rm T} A^0B>$, vanish after
integrating (they have not been written above); moreover, in
$T_B^{pos,pos}$, the integrand does not have a $1/m$ order
contribution. Actually, its leading contribution starts at $(1/m)^3$
which lies beyond the approximation we want to keep.

\begin{figure} 
\centerline{\epsfbox{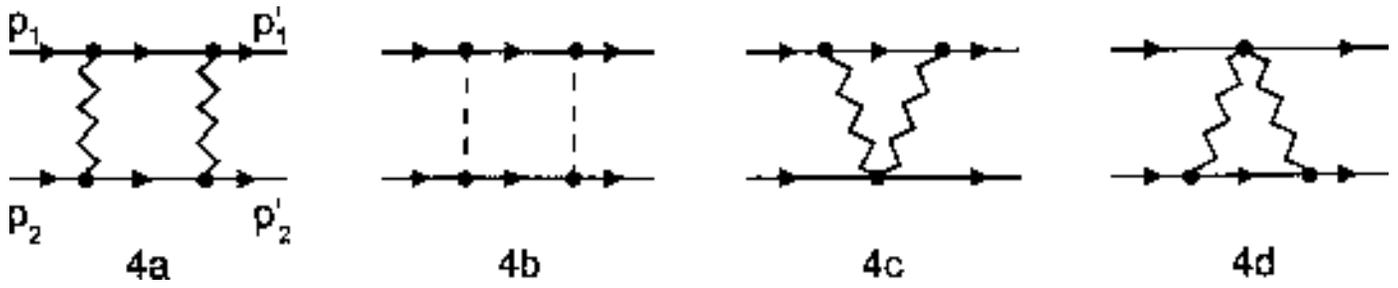}} 
\caption{Non vanishing contributions to the PS
  fermion-fermion scattering in one loop approximation.} \label{fig4}
\end{figure}

\begin{figure} 
\centerline{\epsfbox{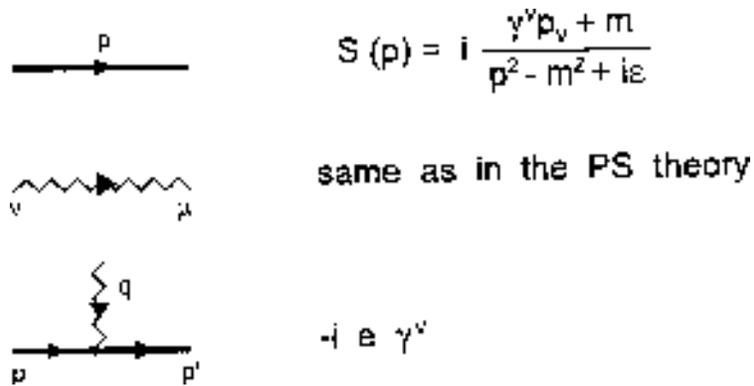}} 
\caption{Feynman rules for the relativistic theory of
a Dirac fermion coupled to a CS field.} \label{fig5} 
\end{figure}

\begin{figure} 
\centerline{\epsfbox{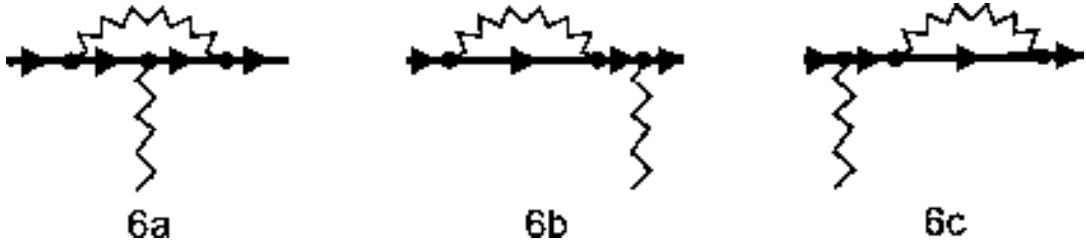}} 
\caption{One loop contributions to the fermion
  anomalous magnetic moment.} \label{fig6} 
\end{figure}

\begin{figure} 
\centerline{\epsfbox{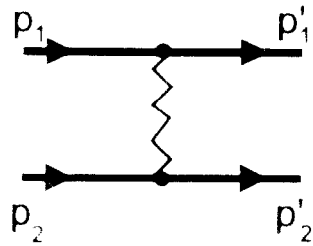}}
 \caption{Graph for the relativistic
  fermion-fermion scattering in the tree approximation.} \label{fig7}
\end{figure}

\begin{figure} 
\centerline{\epsfbox{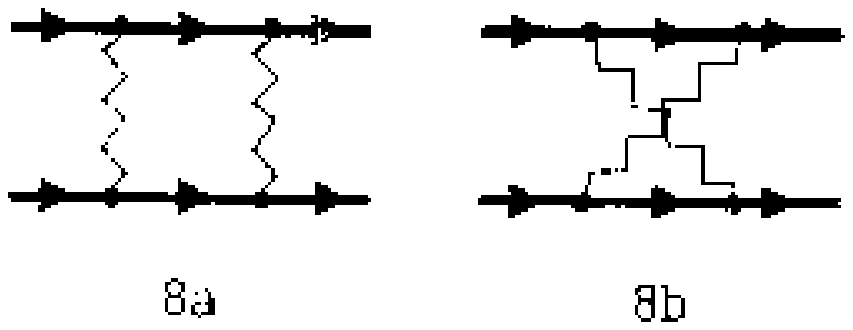}} 
\caption{Graphs contributing to the relativistic
  fermion-fermion scattering in one loop approximation.} \label{fig8}
\end{figure}

The low energy parts of $T_B$ and $T_X$ can be identified with
amplitudes in the PS theory: The first term in the low energy part of
$T_{B}^{el,el}$ corresponds to the diagram \ \ref{fig4}b and the second term to
the diagram\ \ref{fig4}a of the PS theory; the low energy part of $T_X$ exactly
corresponds to the PS result (\ref{17}) coming from the graphs 4c and
4d.

After performing the integrations, one obtains

\begin{eqnarray}
T^{el,el}_B&=& \left [ - \frac{e^4}{4 \pi m\theta^2}  \ln\frac{\Lambda^2}
{\vec q^2} \right ]_{low}+ \left [- \frac{e^4}{4 \pi m\theta^2}\ln\frac{2 m^2}
{\Lambda^2}\right ]_{high}\\
T^{pos,pos}_{B} &=& \left [- \frac{e^4}{4 \pi m\theta^2}\ln2 \right]_{high}\\
T^{pos,el}_{X} &=& \left [  - \frac{e^4}{4 \pi m\theta^2}\ln\frac{\vec q^2}
{\Lambda^2}\right ]_{low} + \left [ - \frac{e^4}{4 \pi m\theta^2}(2 + 
\ln \frac{\Lambda^2}{4 m^2}\right ]_{high}
\end{eqnarray}
\noindent
where $low$ and $high$ refer to the integration intervals $|\vec k| <
\Lambda$ and $|\vec k|>\Lambda$, respectively, of the loop momentum
$\vec k$.
Observe that for each graph the sum of the $high$ and $low$
parts is actually $\Lambda$-independent, as they should be.

If $\Lambda$ is thought as an ultraviolet cutoff ($\Lambda \rightarrow
\infty$), each graph of the nonrelativistic theory ($low$ part of the
relativistic theory) diverges. On the other hand, the corresponding
amplitudes in the relativistic Dirac theory are finite. It is
interesting to see that their high energy parts exactly provide the
counterterms to render the nonrelativistic PS theory finite.

Separately adding the low and the high energy parts of the above
amplitudes we obtain
\begin{equation}
T_B+T_X=  \left [- \frac{e^4}{2 \pi m\theta^2}\right ]_{high}. \label{102}
\end{equation}

The cancellation of the sum of all low energy parts is
connected with the absence of scale anomalies in the PS theory. As
already observed at the end of Sec. 2, in the scalar nonrelativistic
theory it was first noticed in \cite{Lo}.

The high energy result (\ref{102}), which is of the same order in
$1/m$ as the tree approximation (\ref{f15}), is new and could not be
suspected from the PS theory. If we are restricted to the model
(\ref{1}) with fermions of just one flavor and spin, it in fact gives
no contribution after anti--symmetrization of the amplitude. Let us so
enlarge our model (\ref{1}) by assuming that $\psi$ is a $N$ flavor
fermion field. If, analogously, $\phi$ now is also an $N$ flavor PS
fermion, the theory equivalent to the enlarged (\ref{1}) model will be
\begin{eqnarray}
{\cal L}_{eff} &=& \phi^\dagger (i\frac{d\phantom{x}}{dt}- e A^0) \phi
- \frac{1}{2 m} (\vec \nabla\phi - i e \vec A \phi)^\dagger\cdot (\vec
\nabla\phi - i e \vec A \phi) \nonumber \\ &\phantom& +\frac{e}{2m}\,(1
+ e^2/2\pi\theta) \, B
\phi^\dagger \phi + \frac{\theta}{4} \epsilon_{\mu\nu\rho} A^\mu
F^{\nu\rho} - \frac{1}{4}\left (\frac{N\, e^2}{12 \pi m}\right
)F^{\mu\nu}F_{\mu\nu}\nonumber \\
& \phantom& + \frac{e^4}{4 \pi m \theta}(\phi^\dagger \phi)^2.
\end{eqnarray}

Using this new Lagrangian the total fermion-fermion scattering amplitude,
up to one loop, before anti-symmetrization is
\begin{equation}
T= i \frac{e^2}{m \theta} \frac{ {\rm e}^{-i\chi/2}}{\sin(\,\chi/2)}+\frac{Ne^4}{12 \pi m\theta^2}
\end{equation}
For nonidentical fermions the last term survives and provides a
correction to the PS result.

Our study has been restricted to the investigation of the induction of
terms in the effective Lagrangian in leading order of $1/m$. Of
course, a whole series of new terms will be induced in higher orders.

The above Lagrangian summarizes our main results. The low energy limit
of the theory of a CS field minimally coupled to Dirac fermions
differs from the PS theory by an anomalous magnetic moment, a Maxwell
term and a quartic fermionic term, all of the same $1/m$ order. They
are purely quantum field theoretical effects. These results show that
taking the nonrelativistic limit of a classical relativistic
Lagrangian and then quantizing, leads to a different theory than first
quantizing and then taking the nonrelativistic limit.

\newpage
\vspace*{1.0cm}

\vspace{1.0cm}

\vspace{1.0cm}

\begin{references}
\bibitem {f1} We use natural units $(c=\hbar=1)$ and our metric is
$g_{00}=-g_{11}=-g_{22}=1$. The fully antisymmetric tensor
$\epsilon^{\mu \nu \lambda}$ is normalized such that
$\epsilon^{012}=1$ and we define $\epsilon^{ij} \equiv
\epsilon^{0ij}$.  Repeated greek indices sum from 0 to 2, while
repeated Latin indices from the middle of the alphabet sum from 1 to
2.  For the $\gamma$-matrices we adopt the representation
$\gamma^{0}=\sigma^{3},\gamma^{1}=i\sigma^{1},\gamma^{2}=i\sigma^{2}$,
where $\sigma^{i}, i=1,2,3,$ are the Pauli spin matrices.  The
positive and negative energy solutions of the free Dirac equation are
given by

\[
u(\vec{p})\,\,=\,\, \left( \frac{w_p+m}{2\,w_p}
\right)^{\frac {1}
{2}}\,\left[ \begin{array}{c} 1 \\
\frac{p^{2}\,-\,ip^{1}}{w_p\,+\,m} \end{array} \right] \,\,\quad 
v(\vec{p})\,\,=\,\, \left( \frac{w_p+m}{2\,w_p}
\right)^{\frac {1}
{2}}\,\left[ \begin{array}{c} \frac{p^{2}+i p^{1}}{w_p\,+\,m} \\
1\end{array} \right] \,\,.
\]
where $w_p=(m^2+{\vec p}^2)^{1/2}$ and the normalizations were chosen so that
$\bar u u = - \bar v v= m/w_p$.
\end{references}
\end{document}